\documentclass[twocolumn]{emulateapj}

\usepackage{color}

\shorttitle{Multi-spacecraft View of Filament Eruption}
\shortauthors{Gosain et al. 2012}

\begin{document}

\title{A Multi-Spacecraft View of a Giant Filament Eruption during 26/27 September 2009}

\author{Sanjay Gosain\altaffilmark{1}, Brigitte Schmieder\altaffilmark{2}, Guy Artzner\altaffilmark{3}, Sergei Bogachev\altaffilmark{4}, and Tibor T\"{o}r\"{o}k \altaffilmark{5}}
\altaffiltext{1}{National Solar Observatory, 950 N. Cherry Avenue, Tucson 85719, Arizona, USA}
\altaffiltext{2}{LESIA, Observatoire de Paris, CNRS, UPMC, Universit\'{e}
Paris Diderot, 5 place Jules Janssen, 92190 Meudon, France}
\altaffiltext{3}{CNRS UMR 8617, Institut d'astrophysique spatiale (IAS), F 91405 Orsay Cedex, France}
\altaffiltext{4}{Lebedev Physical Institute of Russian Academy of
Science, Leninskij prospekt 53, Moscow 119991, Russia}
\altaffiltext{5}{ Predictive Science, Inc., 9990 Mesa Rim Rd., Suite 170,
San Diego, CA 92121, USA}

\begin{abstract}
We analyze multi-spacecraft observations of a giant filament eruption that occurred during
26 and 27 September 2009. The filament eruption was associated with a relatively slow coronal mass ejection (CME). The filament consisted of a large and a small part, both parts erupted nearly simultaneously. Here we focus on the eruption associated with the larger part of the filament. The STEREO satellites were separated by about
117$^\circ$ during this event, so we additionally used SoHO/EIT
and CORONAS/TESIS observations as a third eye (Earth view) to aid our measurements.
We measure the plane-of-sky trajectory of the filament as seen from STEREO-A and TESIS view-points. Using a simple trigonometric relation, we then use these measurements to estimate the true direction of propagation of the filament which allows us to derive the true $R/R_\odot$-time profile of the filament apex. Furthermore, we develop a new tomographic method that can potentially provide
a more robust three-dimensional reconstruction by exploiting
multiple simultaneous views. We apply this method also to investigate
the 3D evolution of the top part of filament. We expect this method to
be useful when SDO and STEREO observations are combined.
We then analyze the kinematics of the eruptive filament during its rapid acceleration phase by
fitting different functional forms to the height-time data derived from the two methods.
We find that, for both methods, an exponential function fits the rise profile of the filament slightly better than parabolic or cubic functions. Finally, we confront these results with the predictions of theoretical eruption models.

\end{abstract}

\section{Introduction}
Coronal mass ejections (CMEs) are huge expulsions of plasma and magnetic field
from the solar corona into interplanetary space. They are often accompanied by
the eruption of a filament or prominence, which becomes visible as the core of
the CME in coronagraph observations,  and by a flare that occurs almost simultaneously with the eruption.
It is now well accepted that these three phenomena are different
observational manifestations of a more general process, namely a local disruption
of the coronal magnetic field \citep[e.g.,][]{forbes00}. The detailed mechanisms
that trigger and drive such disruptions are, however, still controversial, and a large
number of theoretical models has been put forward in the past decades
\citep[for recent reviews, see][]{amari09,aulanier10,forbes10}.

Early observations indicated that there are two distinct classes of CMEs, namely
fast (or impulsive) ones, originating in active regions and associated with flares,
and slow (or gradual) ones, associated with large prominence eruptions outside
of active regions and no, or no significant, flaring \citep{macqueen83,sheeley99}.
Consequently, it has been suggested that different eruption mechanisms may be
at work in these two types of eruptions. However, the analysis of considerably
larger data sets in the SoHO era revealed a continuous distribution of CME velocities
with a single peak \citep[e.g.,][]{zhang06}, indicating that both fast and slow
CMEs are driven by the same physical mechanism(s). This is supported by the
considerable range of CME kinematics that could be modeled based on a single
physical mechanism \citep{chen03,torok07}, as well as by the fact that large
prominence eruptions outside active regions can produce loops and ribbons that
are morphologically similar to those seen in flare-related CMEs. The majority of
prominence-related CMEs are most likely not associated with flares simply because
the magnetic fields in their source regions are too weak to produce significant
emission in H$\alpha$ and in EUV wavelengths \citep[see, e.g.,][]{forbes00}.

Virtually all theoretical models describe CMEs as coronal magnetic flux ropes that
are anchored in the dense photosphere \citep[see, e.g.,][]{gibson06}, although it
is debated whether a flux rope is present in the corona prior to an eruption or is
formed during the eruption process. The expulsion of a flux rope into interplanetary
space as a CME has been explained by, e.g., the continuous increase of poloidal flux
in the rope due to flux injection from the convection zone into the corona
\citep[e.g.,][]{chen03}, by ideal MHD instabilities like the helical kink instability
\citep{fan05,torok05} or the torus instability \citep{kliem06}, or by the combination
of a ``loss of equilibrium'' of a flux rope and magnetic reconnection occurring in its wake
\citep[e.g.,][]{forbes91}. Other models invoke reconnection from the beginning of the
eruption, as for example the ``tether cutting'' \citep[e.g.,][]{moore01} and
``magnetic breakout" \citep[e.g.,][]{antiochos99} models.

Using observations to support or reject specific models for a particular eruption is
difficult for several reasons. First, many models employ more than one physical
mechanism, resulting in a partial overlap between them \citep[see][]{aulanier10}.
Second, several distinct mechanisms may occur almost simultaneously in an eruption,
in particular in complex events, making it difficult to establish which one is the main
driver of the eruption \citep[e.g.,][]{williams05}. Third, the models predict a very
similar evolution for the main phase of an eruption, i.e., for the evolution after the
impulsive flare phase and initial rapid acceleration of the ejecta. When looking for
clues for possible eruption mechanisms, one therefore often focuses on the early
eruption phase, for example on the morphology and timing of pre-flare H$\alpha$
and EUV brightening \citep[e.g.,][]{chandra11}. For eruptions associated with large
quiescent prominences, as the one studied in this paper, such signatures are, however,
often not available.

Another possibility to obtain information about the mechanisms at work in an
eruption is to study its kinematic properties, in particular the early rise phase
\citep[][]{schrijver08}.
Eruptions typically start with the slow rise of a filament or prominence and/or
overlying loops, at an approximately constant velocity of a few km/s, which is
followed by the rapid acceleration of the ejecta to several 100 km/s. The
acceleration should initially follow some functional dependence, but will then
saturate and decrease afterwards \citep[see, e.g., Fig.\,3 in][]{gallagher03}.
Profiles of the initial acceleration phase, if extractable from measured height-time data
with sufficient coverage, can be compared with predictions of theoretical eruption
models.

Observed rise profiles of the early phase of filament eruptions
and CMEs have been fitted by constant-acceleration curves
\citep[e.g.,][]{gilbert00,kundu04,chifor06}, power-law, $h(t)
\propto t^m$, with $3.0\lesssim m \lesssim 3.7$ \citep[e.g.,][]
{alexander02,schrijver08,liu09}, and exponential functions
\citep{gallagher03,williams05}. As for theoretical models, the
functional dependence of eruption trajectories has not often
been reported yet, and systematic investigations of its
parameter dependence are quite rare so far. Still, some model
predictions can be inferred from the literature. For example,
an exponential early rise is naturally expected if an eruption
is initially driven by an ideal MHD instability
\citep[e.g.,][]{torok04}. A power-law dependence with $m=2.5$
has been found for the trajectory in a two-dimensional (2D)
version of the loss of equilibrium model \citep{priest02}, and
a parabolic rise (i.e., constant acceleration) was reported for
a simulation of the breakout model \citep{lynch04}.
Furthermore, \cite{schrijver08} showed that a velocity
perturbation at the onset of the rapid acceleration of an
eruption can change the resulting functional dependence of the
trajectory (from exponential to near-cubic for the case of the
torus instability they studied), given that this perturbation
is sufficiently large (somewhat larger than typical velocities
observed during the initial slow rise phase of an eruption). We
refer to \cite{schrijver08} for further details.

Prominence eruptions and CMEs have been observed for a long time with various
ground and space-based instrument. For example, the LASCO coronagraphs
\citep{brueckner95} onboard the SoHO spacecraft have been observing thousands
of CMEs in white-light. However, one of the limitations of LASCO, and of other
instruments that are located at, or close to, Earth, is that they can only obtain 2D
observations, projected onto the plane-of-sky (POS). Height-time data of eruptions
are particularly hampered by this, since obtaining the correct radial rise velocities
requires the knowledge of the true 3D trajectories.

STEREO mission \citep{kaiser08} was launched during solar
minimum and therefore initial studies were of quiescent filament eruptions \citep{gosain09,artzner10,gosain10}.  With STEREO 3D reconstructions, it is, in principle, possible to derive the
shape of the prominence, its twist or writhe, and its true
trajectory when it erupts. These properties can be useful for
comparisons with model predictions \citep[e.g.,][]{torok10,kliem12,zuccarello12}.
Other case studies of quiescent filament eruptions observed with STEREO are reviewed in \cite{bemporad11} and \cite{aschwanden11} and those of CMEs in \cite{mierla10}.

The main difficulty in stereoscopic reconstruction arises when
the separation angle between the two STEREO satellites is
large, because then it becomes difficult to identify the same
feature in both views unambiguously \citep{thompson12}. In such cases one has to
use complementary observations from other instruments, such as
SoHO/EIT \citep{delaboudiniere95}, CORONAS/TESIS
\citep{kuzin09} and now SDO/AIA \citep{lemen12}. The addition
of the Earth view to STEREO views makes life easier as it
provides us (a) more than one stereoscopic pair, and (b)
smaller separation angles. Also, in a special circumstance when
a structure (filament/promincence) is not visible in one of the
STEREO satellites (for e.g., remains hidden behind the limb
during initial rising phases), adding an Earth view to one of
the STEREO satellites allows us to make stereoscopic
reconstruction. Some recent examples where three views have been used are \cite{li11} and \cite{feng12}.

In this paper we will present such an example. We present He II 304 \AA\ observations of a large filament
eruption that occurred during 26-27 September 2009. The
observations were taken from the twin STEREO satellites  and
were combined with complementary observations from SoHO/EIT and
CORONAS/TESIS, giving us a third view i.e., the Earth view of
the event. The filament eruption was seen as a limb-event by STEREO-A, EIT, and
TESIS, while it was seen as an on-disk event by STEREO-B.

Based on these multi-spacecraft data, we derive the $R/R_\odot$ or simply height-time
 profile  of the erupting prominence by different methods.
First, we derive independently the POS height-time profiles of the
prominence top as viewed from STEREO-A and TESIS. Second, we apply a trigonometric relation to
 simultaneous STEREO-A and EIT observation of the prominence and estimate the propagation
direction of the filament. Knowing the propagation direction we
can derive true height-time profile of the filament top. Finally, we
apply a three-dimensional stereoscopic reconstruction method
based on Marinus projections to derive the height-time profile, we call
this method ``tomographic method" as it can use simultaneous
multiple views for reconstruction. We then fit these height-time
profiles by functional forms, viz., parabolic, exponential and
cubic and compare the results.

The paper is organized as follows: Section 2 summarizes the
observations. In Sections 3 and 4 we describe the trigonometric
and tomographic methods, respectively results derived from
these methods. In section 5 we discuss the results as well
as the potential of the newly developed tomographic method.

\section{Observations and Method of Data Analysis}

\subsection{He II 304 \AA\ Filament Observations}
\begin{figure}          %%%%%%%%%%%%%%%%%%   FIGURE 1
\begin{center}
\includegraphics[angle=0,scale=.6]{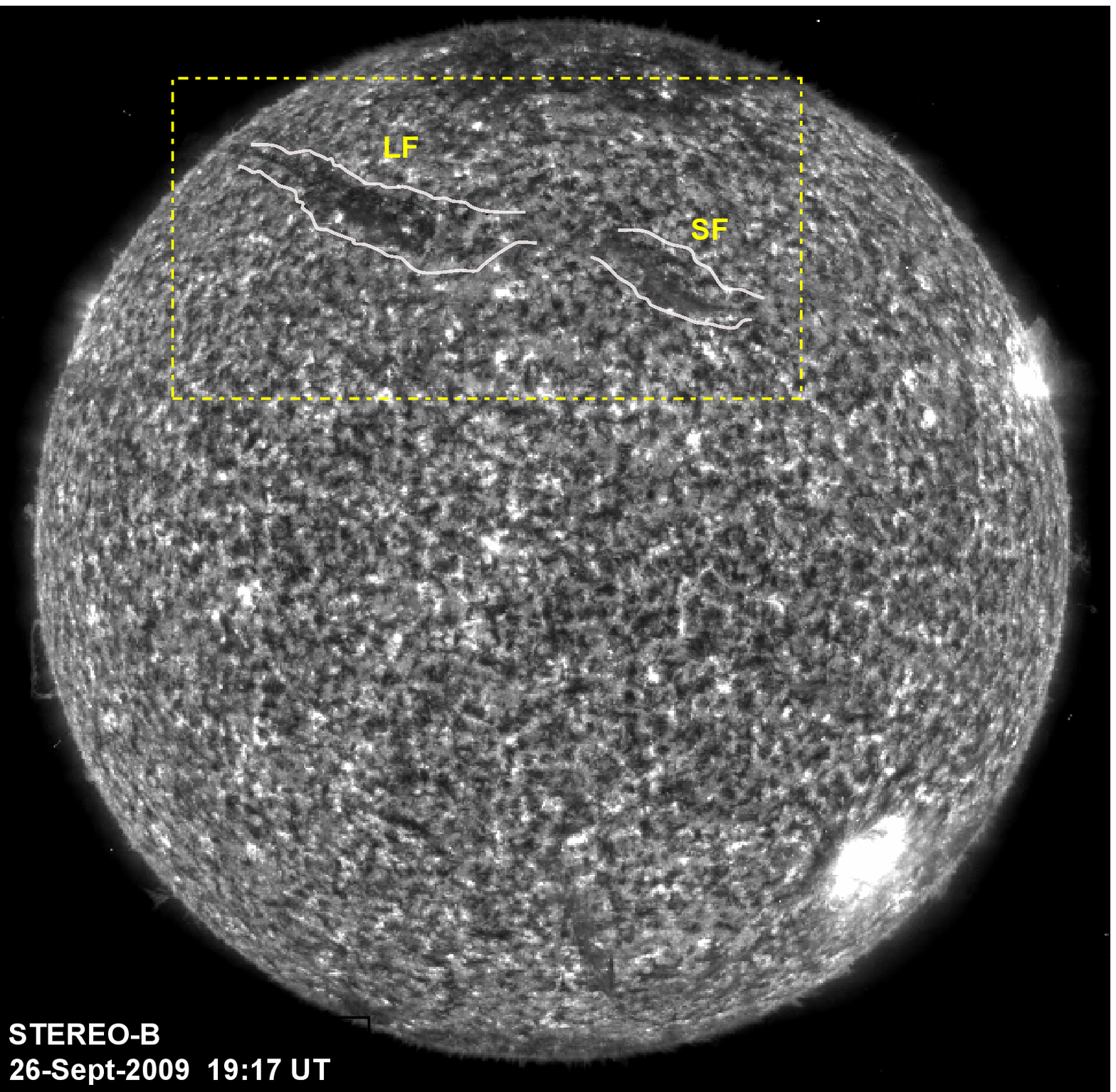}
\includegraphics[angle=0,scale=.5]{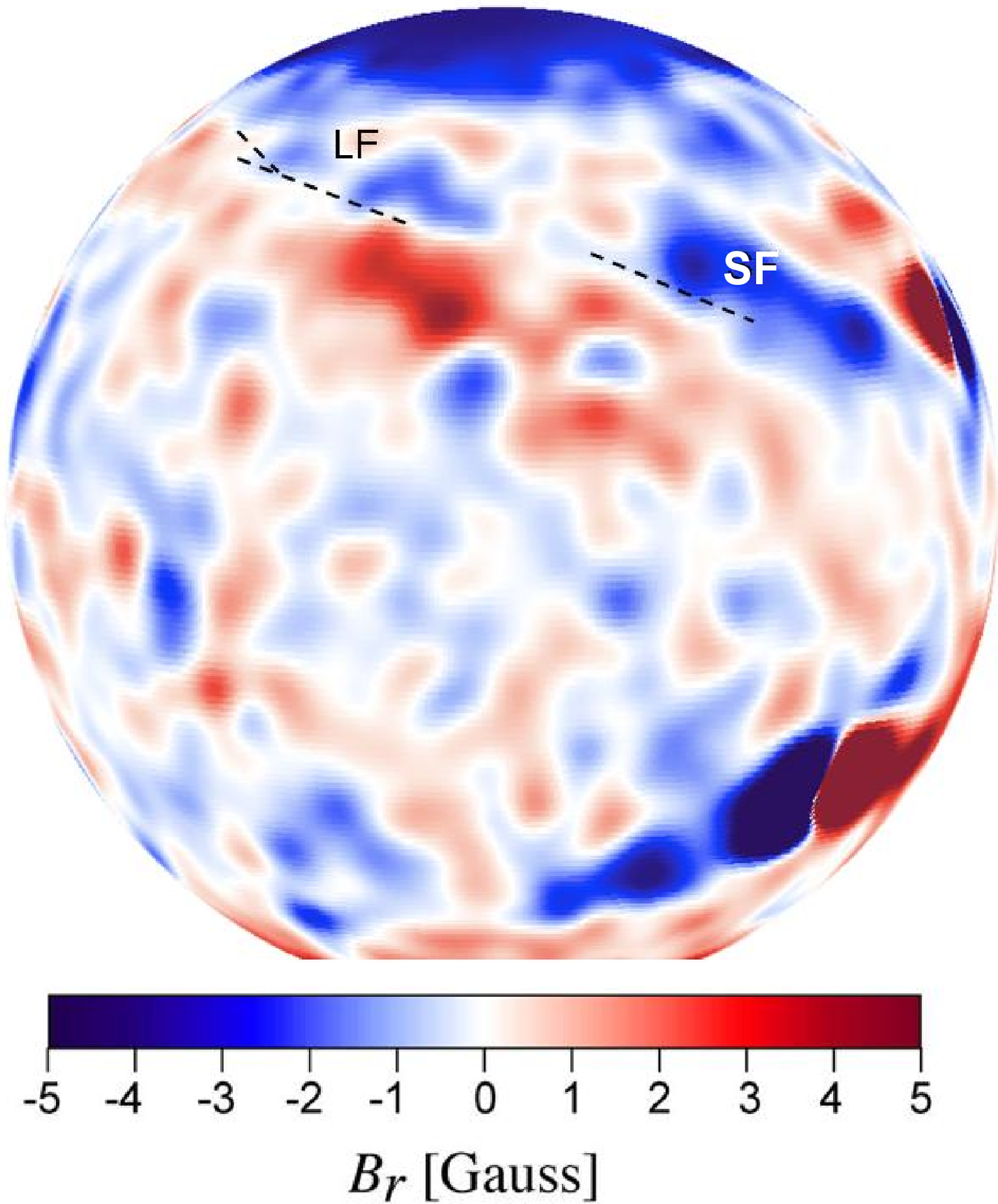}
\end{center}
\caption{
The top panel shows the He 304 \AA\ filtergram observed from STEREO-B at
19:17 UT on 26 September 2009. LF and SF mark the large and the small filament (highlighted inside yellow rectangle, outlined by white line segments for clarity),
respectively. The bottom panel shows a map of the radial magnetic field component,
reconstructed from a synoptic MDI magnetogram, as STEREO-B would have seen it
at the same time (Courtesy of Z. Miki\'c). The dashed lines indicate the polarity
inversion line above which the filaments are located.}
\label{mdi-stereob}
\end{figure}

During  26-27 September 2009 a large filament eruption was observed near the north-east
solar limb. The observations were obtained in the He II 304 \AA\ wavelength at a cadence of ten
minutes by the SECCHI/EUVI instrument aboard the STEREO-A and B satellites.
Figure~\ref{mdi-stereob} (top panel) shows on-disk observations by STEREO-B during the
early phase of the eruption. Two filaments can be seen, a long one (LF) located  at 23-34$^\circ$ N, and a short one (SF) located at 18-32$^\circ$ N.
The bottom panel shows a reconstruction of the photospheric magnetic field distribution as it would have been seen from STEREO-B
(at 19:17 UT on 26 September 2009, using $B_0$ = 5.54$^\circ$  and $CML$ = 164.4$^\circ$,
where $B_0$ and $CML$ are the  Carrington coordinates of the disk center in the STEREO-B view).
It can be seen that both filaments follow the same polarity inversion line (PIL), suggesting that
they were both located within an extended filament channel. The magnetic field on either side
of the PIL was weak (+/-3 Gauss). In the limb view of STEREO-A, LF starts to rise earlier than SF
(see movie 1), while both eruptions seem to occur simultaneously in the disk view of STEREO-B
(movie 2). Since LF is much more prominent than SF, we focus in this paper on the evolution of
LF and refer to \cite{li10} for further details on the eruption of SF. In the following, ``filament"
or ``prominence" therefore refers only to LF.

\begin{figure}   %%%%%%%%%%%%%%%%%%   FIGURE 2
\begin{center}
\includegraphics[angle=0,scale=.85]{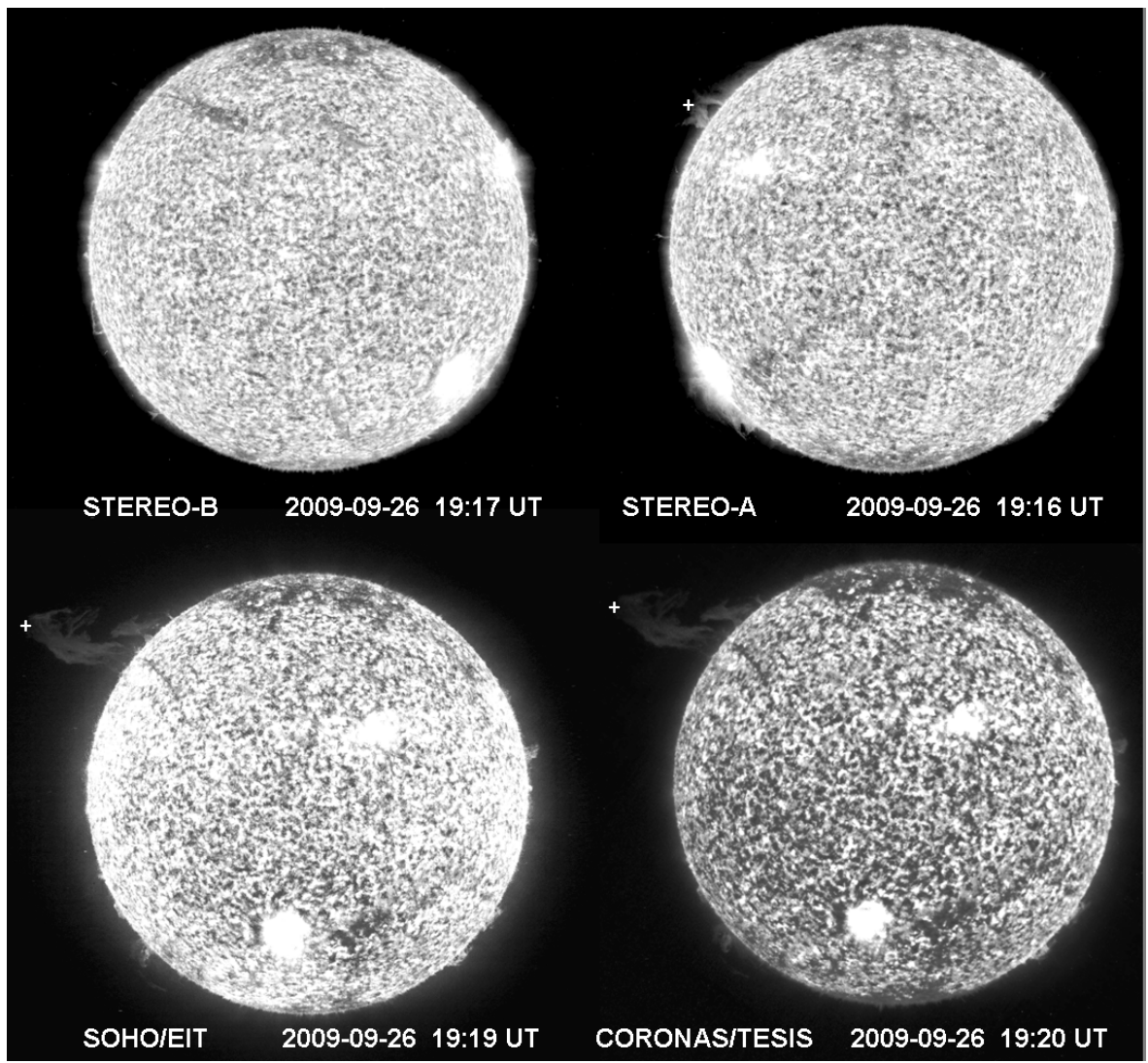}
\end{center}
\caption{
Four different views on the erupting filament in He II 304 \AA\ . The top panel shows
the STEREO-Behind (left) and STEREO-Ahead (right) views. The bottom panel shows
the Earth views by SoHO/EIT (left) and CORONAS/TESIS (right).  }
\label{stereo-tesis}
\end{figure}

During this event the STEREO satellites were separated by about 117$^\circ$ and the main part
of the filament was hidden behind the solar limb in the STEREO-A viewpoint until about 18:20 UT
on 26 September 2009. A 3-D stereoscopic reconstruction in combination with STEREO-A before 18:20 UT was therefore not possible
from either STEREO-B or Earth view (EIT and TESIS).  He II 304 \AA\ images from
SoHO/EIT at a cadence of six hours and CORONAS/TESIS images at a cadence of ten minutes provide the Earth view, which are used for 3D reconstruction with STEREO-A after 18:20 UT Sept 26 using the tomographic method described in section 3.3. The He II 304 \AA\ images of the filament from the four instruments are shown in Figure~\ref{stereo-tesis}.
The Earth-Sun-STEREO-A and Earth-Sun-STEREO-B angles on 26 September 2009 were 61$^\circ$ and 56$^\circ$, respectively.
The observations of the filament top from two different vantage points (TESIS and STEREO-A) allows us to observe the POS evolution  in a piecewise continuous manner from TESIS (00:00 to 22:00 UT) and STEREO-A (19:10 to 23:10 UT), as shown in Figure~\ref{height-plane-sky} (top left panel). A simple trigonometric method, described in section 3.2, is used to triangulate the true propagation direction of the filament apex. Knowing this angle the POS  height-time profiles are corrected to derive the true  height-time profiles as shown  in the  Figure~\ref{height-plane-sky} (bottom left panel).

\begin{figure}      %%%%%%%%%%%%%%%%%%   FIGURE 3
\begin{center}
\includegraphics[angle=0,scale=.65]{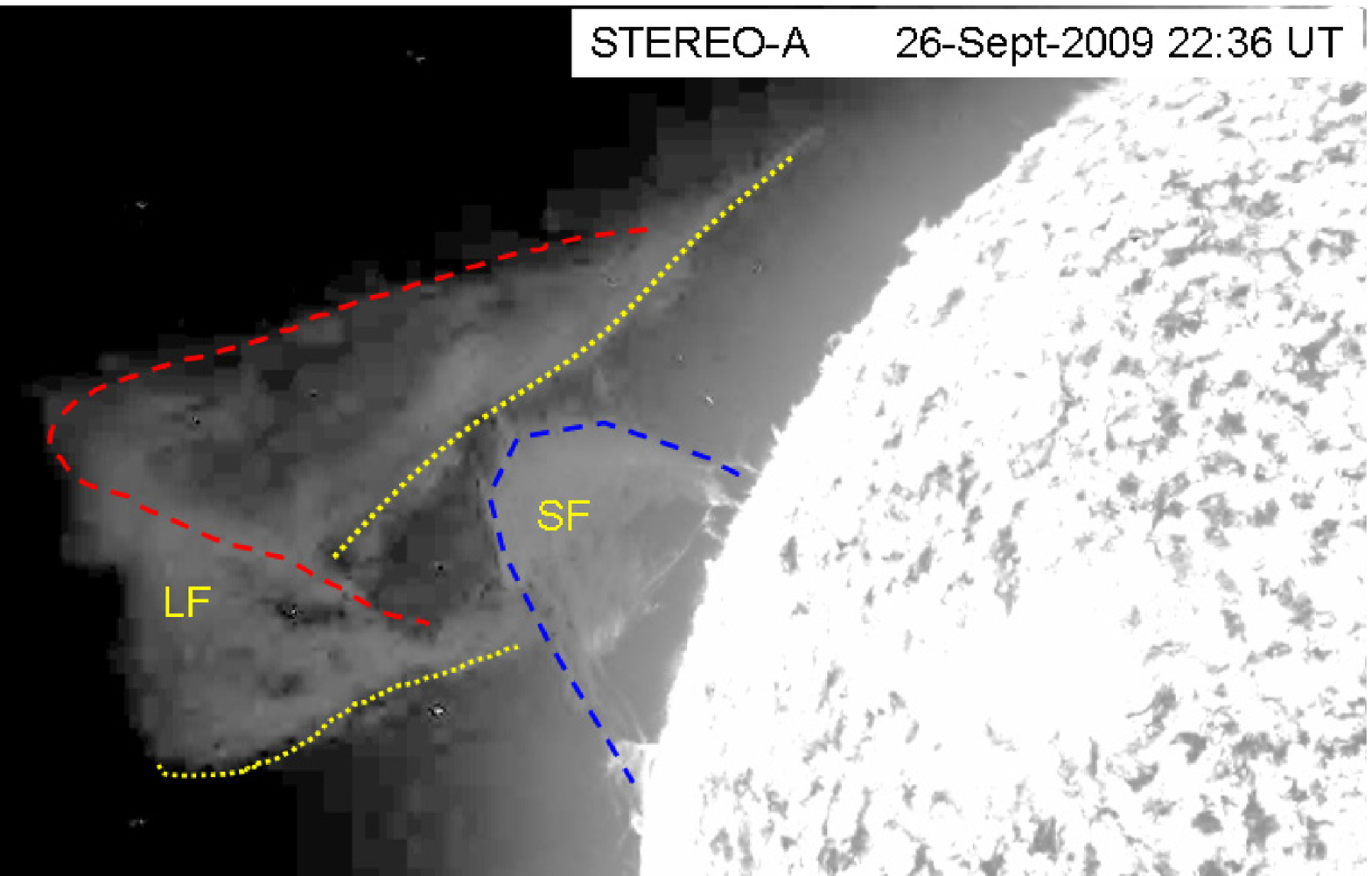}
\includegraphics[angle=0,scale=.62]{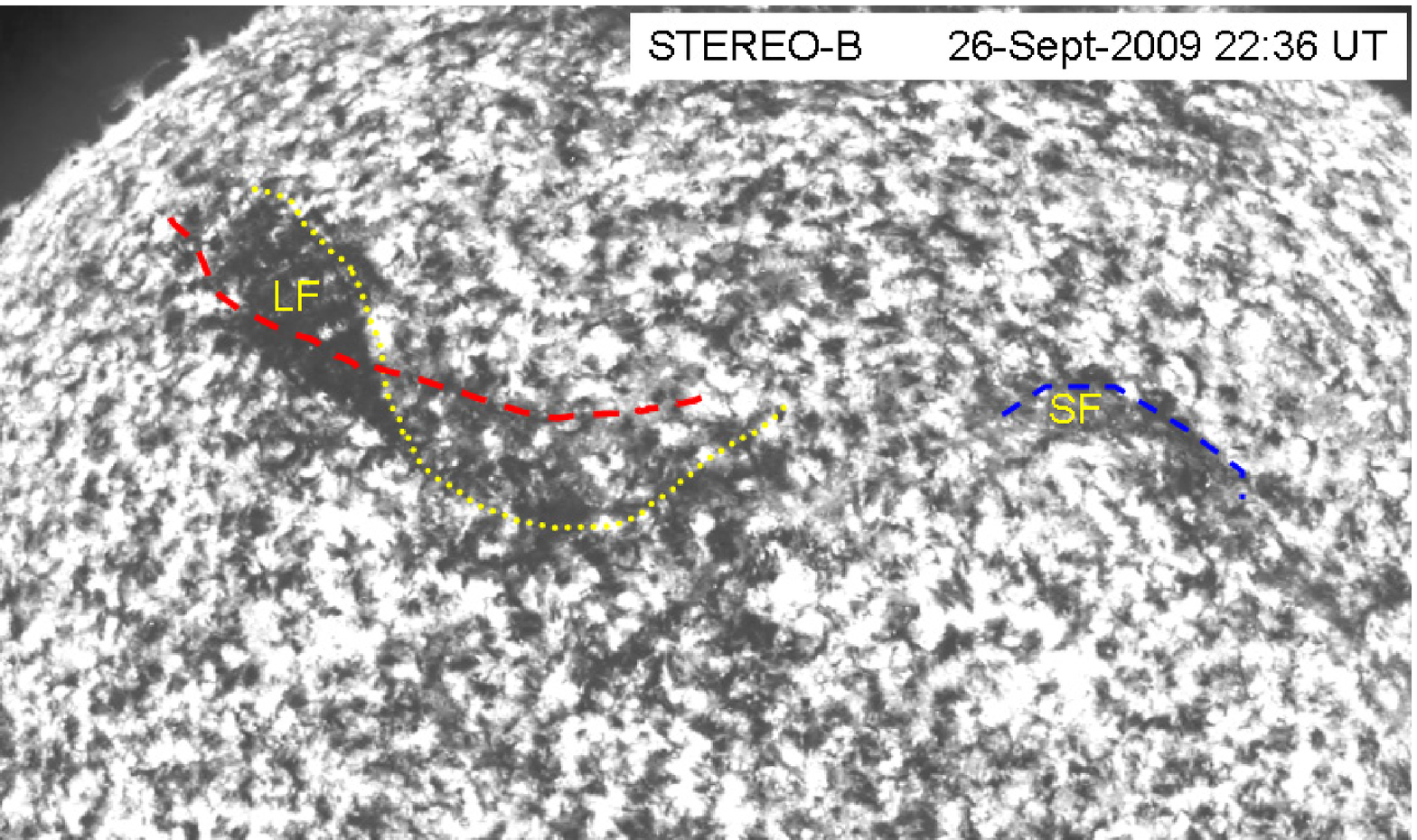}
\end{center}
\caption{ STEREO-A (top) and B (bottom) images of the eruptive filament (LF and SF) are shown. The edges of the
filament sheet (LF) are outlined by dashed (red) and dotted (yellow)
lines. The sheet appears to be twisted along its legs.  }
\label{twist}
\end{figure}

The filament LF (as seen from STEREO-A; Figure~\ref{twist}) suggests a sheet-like morphology.
The stereoscopic reconstruction by \citet{li10} (see their Figure 2) also infers a sheet like structure.
We outline its apparent edges by dashed (red) and dotted (yellow) lines in STEREO-A and B views. A careful inspection of the legs of the prominence in STEREO-A images suggests a twisted morphology of the legs. However, a quantification of the twist of the filament sheet is not possible in the present case.

\subsection{CME observations}
 STEREO-A observed the CME associated with the filament eruption on 26/27 September 2009 with its coronagraphs,
COR1 (1.4 to 4 R$_\odot$) and COR2 (2.5 to 15 R$_\odot$).  The CME was not seen by STEREO-B since it was directed
towards it and was perhaps too faint to be seen as a halo CME. The LASCO/C2 (1.5 to 6 R$_\odot$) and C3 (3.7 to 30 R$_\odot$) coronagraphs also observed the CME. The time of arrival of CME in C2 was 23:06 UT on Sept 26, in C3 was 14:18 UT on Sept 27, and in HI1 was on 21:29 UT on Sept 27.

The  propagation angle of the CME leading edge is derived using same procedure as described in section 3.2 and applied to filament apex. The projection correction to the POS  height-time profiles of the CME (top right panel of Figure~\ref{height-plane-sky}) is then applied to derive the true  height-time profiles (bottom right panel of Figure~\ref{height-plane-sky}).

\section{Height-Time Profile of Erupting Filament and CME}
\subsection{Plane-of-sky Measurements}
 We make plane-of-sky (POS) measurements of the filament apex (marked by `+' symbol in the panels of Figure~\ref{stereo-tesis}) using TESIS and STEREO-A observations. The $R_{POS}/R_\odot$ profile of the filament apex measured from the two views (from STEREO-A and TESIS) is plotted in the top left panel of the Figure~\ref{height-plane-sky}. Similarly, the top right panel of the Figure~\ref{height-plane-sky} shows the $R_{POS}/R_\odot$ profile of the leading edge of the CME measured with STEREO-A and LASCO coronagraph observations. The difference in the $R_{POS}/R_\odot$ profiles of the filament apex and CME leading edge as seen by different satellites is apparent, since the measurements correspond to two different vantage points.

\subsection{Estimation of the True Height-Time Profile using Simple Triangulation}
 The POS  measurements, $R_{POS}/R_\odot$, of the filament apex and the CME leading edge, described above, can be corrected
for projection effect if we know the angle between the real trajectory of the
erupting feature and the POS in the observer's frame of reference. The true
$R/R_\odot$ profile is related to the $R_{POS}/R_\odot$ profile, measured in the POS, by
$(R/R_\odot)\mathrm{cos}~\theta=R_{POS}/R_\odot$, where $\theta$ is the angle between
the real trajectory and the POS, referred to as propagation angle henceforth.

Here we apply a simple trigonometric relation, using image pairs from STEREO-A
and TESIS/EIT (henceforth, Earth View or EV), to estimate the propagation angle. We then use this information
to derive the true $R/R_\odot$ profile of the erupting filament and the
CME.
A simple assumption made here is that the propagation angle remains
unchanged during the time of the measurements. We verified this assumption by computing the propagation angle using STEREO-A and TESIS pair at later times and found that the angle remains the same (see Table 1).

\begin{figure}   %%%%%%%%%%%%%%%%%%   FIGURE 4
\begin{center}
\includegraphics[angle=0,scale=.65]{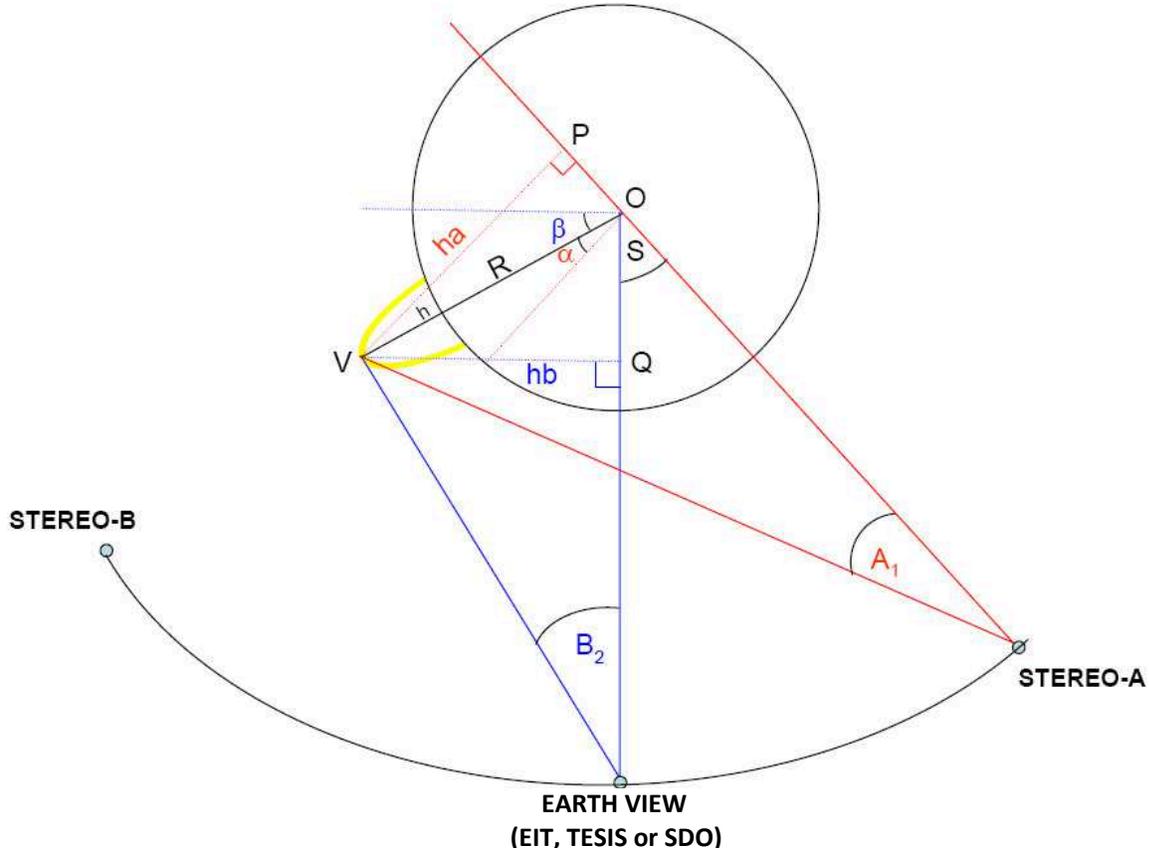}
\end{center}
\caption{
An illustration of the de-projection method. The coronal structure is represented
by a yellow loop extending above the solar limb. The top part of this loop is at an
angle $\alpha$ ($\beta$) to the plane-of-sky in the reference frame of STEREO-A
(Earth View, EV). $S$ is the separation angle between STEREO-A and EV. The
segments $ha$ and $hb$ are the projected distances $(R+h)\,\mathrm{cos}(\beta)$ and
$(R+h)\,\mathrm{cos}(\alpha)$, respectively. $S$ is related to $\alpha$ and $\beta$ by
$S=\alpha+\beta$ (if the loop is seen in front of the limb in one view and behind
the limb in the other) or by $S=|\alpha-\beta|$ (if the loop is seen on the same
side of the limb, i.e., either in front of or behind the limb, in both views).
If $ha$, $hb$, and $S$ are known, $\alpha$ and $\beta$ can be determined.}
\label{triangulation}
\end{figure}

We explain the trigonometric procedure here briefly. The illustration in Figure
\ref{triangulation} shows the geometric setting of the two STEREO satellites and the EV with respect to the filament. The projected height of the top part of
the filament is $h_a$ and $h_b$ in the POS of STEREO-A and EV,
respectively. $S$ is the separation angle between STEREO-A and EV (61 degrees),
and $\alpha$ and $\beta$ are the angles that the top of the filament apex (trajectory) makes with
respect to the POS. We obtain the angles using the relations $S=\alpha+\beta$
and $h_a/h_b=\mathrm{cos}(\alpha)/\mathrm{cos}(S-\alpha)$. Knowing these angles, we can apply
corrections to the POS heights $h_a$ and $h_b$ to obtain the true height
$h_{True}=h_a/\mathrm{cos}(\alpha)=h_b/\mathrm{cos}(\beta)$. %It is assumed that $\alpha$ and
%$\beta$ do not change during the time period of the measurements.

\begin{figure}   %%%%%%%%%%%%%%%%%%   FIGURE 5
\begin{center}
\includegraphics[angle=0,scale=.5]{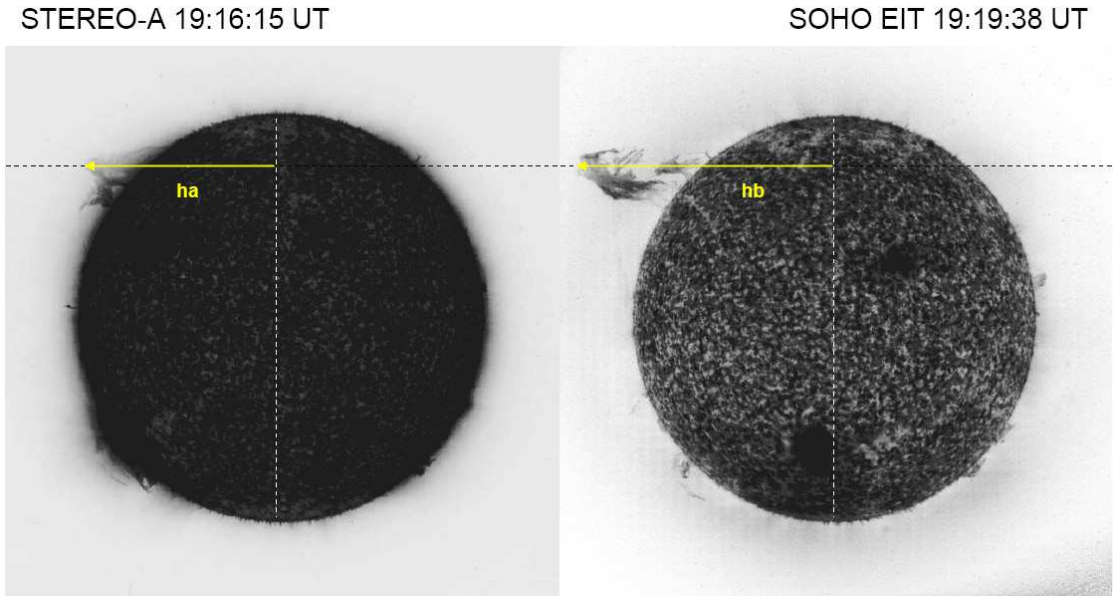}
\end{center}
\caption{
Near-simultaneous STEREO-A and SoHO/EIT filtergrams in epipolar geometry.
The segments $h_a$ and $h_b$ are measured from the apex of the
filament. The separation angle between Earth and STEREO-A is $S=\alpha+\beta=61^\circ$. }
\label{stereoa-eit}
\end{figure}

\begin{figure}      %%%%%%%%%%%%%%%%%%   FIGURE 6
\begin{center}
\includegraphics[angle=0,scale=0.6]{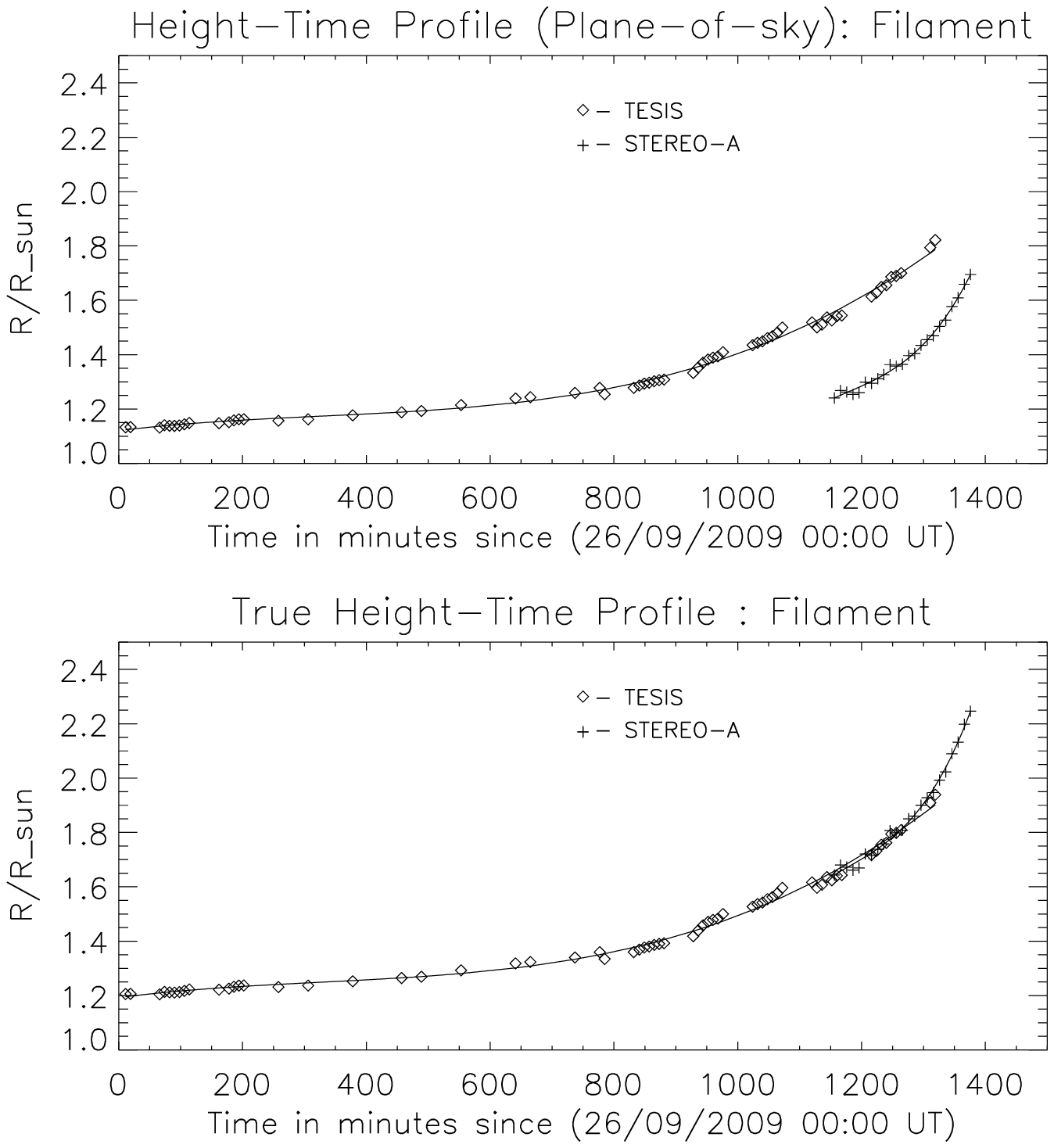}\includegraphics[angle=0,scale=0.6]{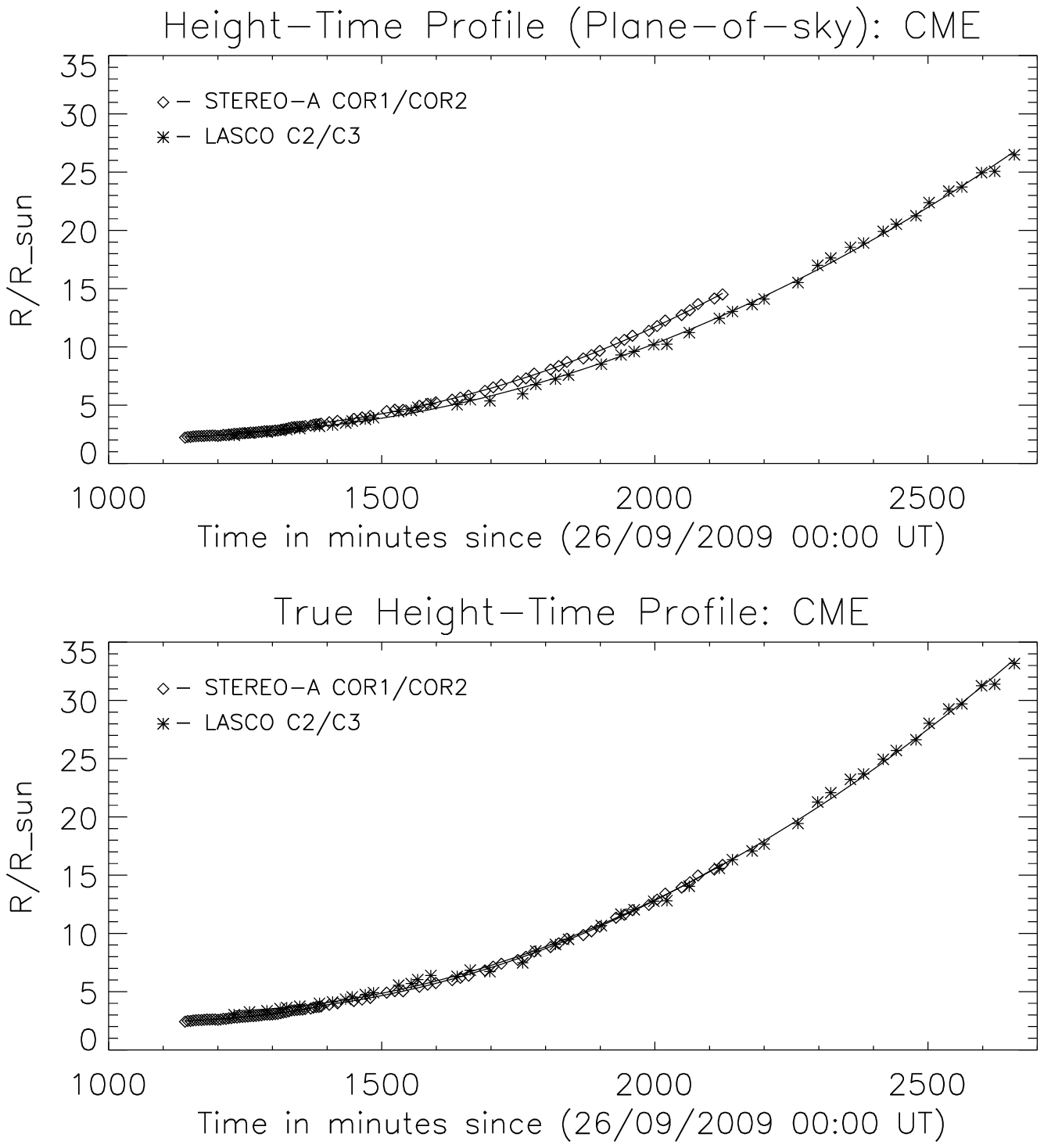}
\end{center}
\caption{
The top panels show the POS  height-time profiles of the filament apex (left) and CME leading edge (right) as seen from different viewpoints (see inset for the observing satellites). The lower panels show the true  height-time profiles derived by using the simple trigonometric method (Section3.2). }
\label{height-plane-sky}
\end{figure}

Figure~\ref{stereoa-eit} shows an example of two stereoscopic image pairs, i.e.,
STEREO-A (left panel) and EIT (right panel), observed almost simultaneously.
The two images are in epipolar view. The segments $h_a$ and $h_b$ measure
the top part of the filament as viewed from two vantage points. Knowing $h_a$,
$h_b$, and $S$, we determine $\alpha$ and $\beta$ to be 41 and 20 degrees,
respectively.  Since only one stereoscopic pair is available between SOHO/EIT and STEREO-A, we make use of TESIS data to make pairs with STEREO-A. The observations of STEREO-A and TESIS are not synchronized in time and both instruments follow a different time cadence (see Table 1). Assuming that the filament did not evolve significantly within the small time differences, we could make 8 near-simultaneous stereoscopic pairs of TESIS and STEREO-A. The timings of these pairs and the value of angles $\alpha$ and $\beta$ deduced using these pairs is given in Table 1. It may be noticed that during the observed time interval the propagation angle does not change and therefore it is possible to correct the observed POS height-time profiles for propagation angle using single value of $\alpha$ and $\beta$. In section 3.3.3 we show that these values are consistent with other 3D reconstruction methods. The corrected height-time profile of the filament apex
  is shown in the lower left panel of the Figure~\ref{height-plane-sky}.

  Similarly, applying this method to the CME leading edge we deduce angles $\alpha$ and $\beta$ to be 36 and 25 degrees, respectively. The true height-time profile of CME leading edge after correcting for these angles is shown in the lower right panel of the Figure~\ref{height-plane-sky}. It is interesting to note that the direction of propagation of the filament apex and the CME leading edge differs by about 5 degrees. The filament typically forms a core in the three part CME structure. However, since the CME leading edge is more extended, i.e., the front surface  of a tear-drop shaped bubble in which filament forms a trailing part, the difference of 5 degrees is small considering the large angular extent of the filament and the associated CME. Another, interesting point about this method is that the two curves merge into one (as seen in the combined curves in lower panels) only for a unique pair of $\alpha$ and $\beta$ angles, where the sum of the two angles ($S=\alpha+\beta$, in this case equal to separation between Earth and STEREO-A, i.e., 61.5 degrees) is well constrained by the known separation angle between the two vantage points. For any other pair of these angles the two curves did not merge into one. Thus, just by knowing the separation angle between two vantage points and the respective POS  height-time profiles, one can iteratively adjust the angles (in fact only one of the angles, as the two angles $\alpha$ and $\beta$  are simply, $\alpha$ and $S-\alpha$), until the two curves merge as one. This procedure also gives the same solution for $\alpha$ and $\beta$. These true  height-time profiles are then used for deriving the velocity and acceleration profiles of the filament and CME, which is described in the following sections.

\subsubsection{Estimating the duration of Rapid acceleration phase:} In this section, we use the true  height-time curves, shown in the lower panels of Figure~\ref{height-plane-sky}
to derive the velocity, acceleration and jerk (rate of change of acceleration, following Schrijver et al. (2008)) profile of the filament and CME. The latter can then be
compared to the predictions of theoretical eruption models described in the Introduction. Since
the acceleration and jerk are the higher order time derivative of the trajectory, errors in the measured data amplify
strongly, so one is typically forced to smooth the data before calculating acceleration curves, for
example using spline smoothing \citep{vrsnak07}. Here we will use a different approach: we first fit a fourth order polynomial of the form
$H(t)=a+bt+ct^2+dt^3+et^4$ to the  height-time data. We then use this smooth
curve to obtain the velocity, acceleration and jerk profiles. These profiles are shown in
Figure \ref{rapid-accel-profile} for the filament and the CME leading edge in the left
and right columns, respectively.

It is to be noted that the underlying physical mechanism responsible for the eruption determines the functional form of the height-time profile only initially, i.e., during the phase when acceleration is growing, but not yet saturating. Once the acceleration starts to saturate, the functional form is changing.
From the Figure\ref{rapid-accel-profile} we notice that the acceleration profiles as estimated from the 4th order polynomial are quite different
for the filament and the CME.  While the acceleration of the CME leading edge is higher than that of the filament, the rate of change of acceleration, i.e, the value of jerk for the CME is declining, in contrast to the filament. This suggests that during the time interval of the CME
data the increase of the acceleration of the leading edge is slowing down, while the growth of the acceleration of the filament is still increasing. We therefore restrict our fits of different functional forms to the filament only and not to the CME.

\subsubsection{Fitting functional forms to the Filament rapid acceleration phase:}
Before we fit the functional forms  we make an estimation of the optimum time interval which corresponds to rapid
acceleration phase of the filament. To get the first
estimation we took the start time where the acceleration starts to grow from zero and the end time as the last data point.
We then fine-tune our estimation of the time interval of the rapid acceleration phase
by varying their start and end times and observing the resulting quality of the
overall fits of all three functional forms. The interval leading to an overall best fit quality for all functional forms is marked by vertical
dashed lines in Figure \ref{rapid-accel-profile}.

Within this rapid acceleration interval we then find the best fit functional form as described below.
We fit the three different functional forms: (i) parabolic, $H(t)=a+bt+ct^2$, (ii) exponential,
$H(t)=ae^{bt}+c$, and (iii) cubic, $H(t)=a+bt+ct^3$ to the  height-time profile during the rapid acceleration phase. The left panel of the Figure~\ref{artzner-fit} shows these fits.
The reduced chi-square $\chi_\nu$ values are shown at the top left corner
of each panel. The weights that we apply to the data points for fitting are
taken to be $W=1/\sigma^2$, where $\sigma$ is the standard deviation of
the measurement error. We assume a Gaussian distribution for the latter.
The pixel size of STEREO/EUVI is about 2 Mm.
For the filament (or prominence), which is typically quite diffuse in
He II 304 \AA\ images, we consider 3 pixels, i.e. 6 Mm as the 1-$\sigma$ error.
The actual errors may be somewhat different. However, while using
different values of 1-$\sigma$ will lead to different values of $\chi_\nu$,
the relative values of $\chi_\nu$ between different functional forms will
remain the same. The fits of the three functional forms shown in the left panels of Figure~\ref{artzner-fit}, clearly favor an exponential rise of the filament, with a relatively better value of $\chi_\nu$.

\subsection{The 3-D reconstruction by using Marinus projection}
\subsubsection{The Method}
 Here  we describe a new tomographic method for the 3D reconstruction. We used simultaneous views of the filament from STEREO-A and B and TESIS in He \textsc{II} 304 \AA\ wavelength. The essence of the method is as follows.

A continuum intensity image of the sun, $I(x,y)$, can be easily projected into heliographic coordinates $I(l,b)$. This projection is also known as equidistant cylindrical or Marinus projection. Since the continuum intensity $I(x,y)$ corresponds to the solar photosphere, each point $(x,y)$ on the intensity image can be associated with heliographic coordinates $(l,b)$, assuming a spherical sun with radius, $R=R_\odot$. A common feature on the solar disk, such as a sunspot should then correspond to the same Carrington latitude-longitude, no matter what the viewing angle of the sun is. However, for  coronal images like in He II 304 \AA\, the intensity features corresponding to filaments, spicules etc. do not lie on the same sphere but are elevated structures in 3D. Thus, a common feature like a filament or coronal loop will correspond to location $(l_1,b_1)$ and $(l_2,b_2)$ in heliographic projection of the coronal images obtained from different viewing points 1 and 2, respectively (when $R=R_\odot$ is assumed). Conversely, if the heliographic projection is attempted assuming the Sun to be a sphere of radius larger than one solar radius and a correct radius of the sphere is assumed (equal to the altitude of the feature) then the we should get $l_1=l_2$ and $b_1=b_2$ for the common feature.

Thus, generating the  generalized Carrington maps for different assumed radii of the spherical grid,  using a 5 Mm step from $R$=700 Mm to $R$=1500 Mm, and comparing the Carrington coordinates (latitude-longitude) of a recognizable common feature, such as filament apex, in the three Carrington maps (one for each viewing angle) until they all agree gives us a solution for the 3-D coordinates of the feature.  We found that this step size of 5 Mm gives an optimum choice to arrive at the best agreement for the generalized coordinates of a recognizable feature. Thus the generic accuracy of the method can be assumed to be about 5 Mm.

 We geometrically consider both intersections of the line of sight with the reference sphere. When the radius of the reference sphere is equal to the chromospheric radius, we take into account the single point located physically  in front of the POS. When the radius of the reference sphere is greater than the chromospheric radius, we have in principle to take into account both intersection points located respectively in front of and behind the POS. That is why in the top panel of Figure 8 the prominence, projected on the far side of the reference sphere, behind the POS of STEREO-A, appears as reversed from right to left with respect to the direct view in Figure 2. In addition to that, the far side of the solar disk appears as a dark, missing disk in top panel of the Figure~\ref{marinus}.

\subsubsection{Advantages and Limitations of the Method}
It is well known that  all stereoscopic reconstruction methods are limited by the ambiguity in recognizing a common feature in different views. %This problem is alleviated in the case of optically thin and highly dynamic features like erupting prominence. However, during initial phase of giant filament eruptions  the structure is generally optically thick, as in the present case, and so the reconstruction methods can be applied reliably to structures like leading edge of the filament.
Further, a common limitation that arises with any 3D stereoscopic reconstruction technique is when the apex point from two viewing angles may be different. Such situations would lead to a systematic error in the reconstructed 3D coordinates. However, we expect such errors to be less severe in our case because (a) the filament studied here has a large extension in longitude which is rising globally as a whole, so height-time profile of several neighboring points along the filament will be similar, and (b) using combination of STEREO-A and TESIS (separation angle 61$^\circ$) as compared to STEREO-A and STEREO-B (separation angle 117$^\circ$), we reduce the errors. Although, such systematic errors cannot be avoided, the time derivative of measured altitude and hence the derived velocity and acceleration should not be affected severely as long as the systematic error remains similar in magnitude. Therefore, for studying the kinematic evolution of erupting prominence such reconstruction methods may still be applicable, with aforesaid limitations.

The Figure~\ref{marinus}, shows  selected parts of the three  generalized Carrington maps corresponding to STEREO-A, TESIS and STEREO-B views, generated assuming radii of the spherical grid to be 1245 Mm, i.e., 545 Mm above the solar surface. At the choice of this radius the common feature, i.e., the filament apex marked by a square box, corresponds to the same Carrington latitude-longitude coordinates in the different views. In principle, two stereoscopic views are sufficient for the application of this method. However, adding more views increases redundancy  (for example in present case more emphasis is given to TESIS and STEREO-A for constraining reconstruction) and therefore  may add to its robustness. In future, we plan to apply this method to the events observed simultaneously by the two STEREO/EUVI instruments along with the high-resolution SDO/AIA observations.

\subsubsection{Comparison with SCC\underline{ }MEASURE and Simple Triangulation Method}
 The 3D reconstruction of the filament studied in this paper, was also carried out by \cite{li10} using SCC\underline{ }MEASURE  procedure (developed by W. Thompson). They used STEREO-A and STEREO-B pair for their reconstruction. Further, they reconstructed many (12) points along the filament body (their Figure 5). Their points 6,7 and 8 correspond to the top of the filament and one can notice that the height evolution of these points (their Figure 6(a)) is quite similar to each other (within $\pm$6Mm), though the points are separated spatially, this is due to the large scale uniform evolution of the filament. For comparison, we overlay the altitude data points of location 7, as reconstructed by \cite{li10} in our height-time plot shown with red colored symbols in the top-right panel of Figure 9. The reconstruction from two independent methods agrees quite well, considering the general scatter in the reconstructed coordinates.

On the other hand, a poorer match is expected between the true 3D reconstruction methods and the simple triangulation method since the latter only estimates the propagation angle and not the 3D coordinates of the filament. The height derived from simple triangulation method shows a systematic offset with respect to the true height derived from 3D reconstruction methods. Apart from the systematic offset, the profile of the derived speed and acceleration should however remain unaffected, as these depend upon the shape of the curve. This is evidenced in a similar fit quality of the height-time profile by both the methods to different functional forms (see section 3.3.4). Also, the method is straightforward and relies in tracking a common feature in the images taken from same vantage point albeit at different times. The natural advantage is that it is easy to track a common feature in time if the time difference between two images is not very large. 
%For large scale structures like CME leading edge, and for radially propagating CMEs this method may provide a good estimate of true CME acceleration.

Further, it may be noted from Table 1 that the angles $\alpha$ and $\beta$ are not changing significantly.  In the 3D reconstruction by \citep{li10} (their figure 4, right panels) it can be seen that a propagation angle of $\sim$20$^\circ$ in front of the East solar limb is deduced and is not changing significantly. Also, the Carrington longitude of the filament apex reconstructed using Marinus method (Table 1.) shows a small variation in longitude of $\sim4^\circ$, while the mean value of the longitude, $\sim150^\circ$, corresponds to an angle of $\sim20^\circ$ in front of the East solar limb, in agreement with \cite{li10} and angle $\beta$ from simple triangulation method.

\subsubsection{Rapid acceleration phase and its functional form:}
We use this tomographic 3-D reconstruction method based on Marinus projection to obtain the 3-D trajectory of the filament apex. The rise of the altitude of the filament apex is fitted for different functional forms. The time interval of rapid acceleration phase is taken to be the same as estimated in section 3.2.1.  The  height-time curve and the fitted parabolic, exponential and cubic functions to it are shown in the right panel of Figure~\ref{artzner-fit} from top to bottom, respectively. The reduced chi-square $\chi_\nu$ values are shown at the top left corner of each panel. It is found that an exponential form fits the observations relatively better as compared to the other functions. The exponential function was also found to fit the rapid acceleration phase relatively better than other functions in section 3.2.2, where simple triangulation method was used. This is shown in the left panel of Figure~\ref{artzner-fit}.

\section{Discussion and Conclusions}
 In this paper we analyzed the observations of a large erupting quiescent filament which was observed from three vantage points by STEREO-A, B and the EV (SoHO/EIT and TESIS). The filament rose slowly for several hours before accelerating rapidly and erupting in two parts, a large and a small filament. We analyzed the kinematics of the large filament, whose true trajectory was derived by two methods: one simple triangulation method and another newly introduced tomography method. The new tomographic method can potentially take advantage
of simultaneous observations from multiple vantage points to constrain the reconstructions better. After deriving the true trajectory by the two methods we fitted the  height-time curves with different functional forms and compared the results with predictions of theoretical eruption models.

\begin{figure}      %%%%%%%%%%%%%%%%%%   FIGURE 7
\begin{center}
\includegraphics[angle=0,scale=0.95]{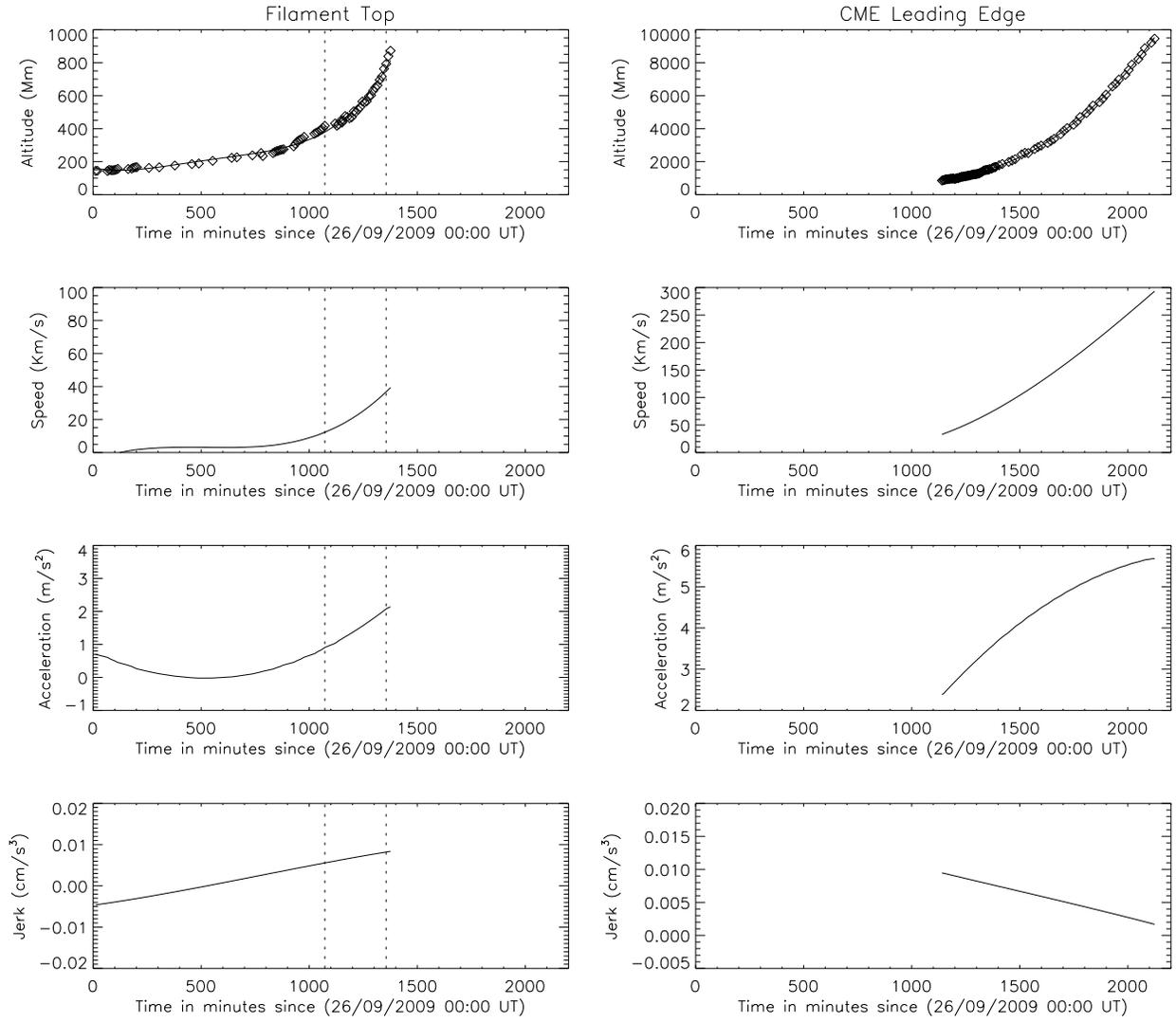}
\end{center}
\caption{
The left (right) panels show, from top to bottom,  the height,
speed, acceleration and jerk (rate of change of acceleration)   profiles of the filament apex (CME
leading edge), respectively. In the top panels, the data points
are shown by `+' marks and the solid line corresponds to a
fourth order polynomial fit ($H(t)=a+bt+ct^2+dt^3+et^4$) to the data.
The fitted profile $H(t)$ is used to derive the speed,
acceleration and jerk curves in the subsequent panels. The two vertical dotted lines in left panels
correspond to the estimated duration of the rapid acceleration phase of the filament, which is fitted with different functional forms in the Figure~\ref{artzner-fit}. }
\label{rapid-accel-profile}
\end{figure}

\begin{figure}      %%%%%%%%%%%%%%%%%%   FIGURE 8
\begin{center}
\includegraphics[angle=0,scale=.95]{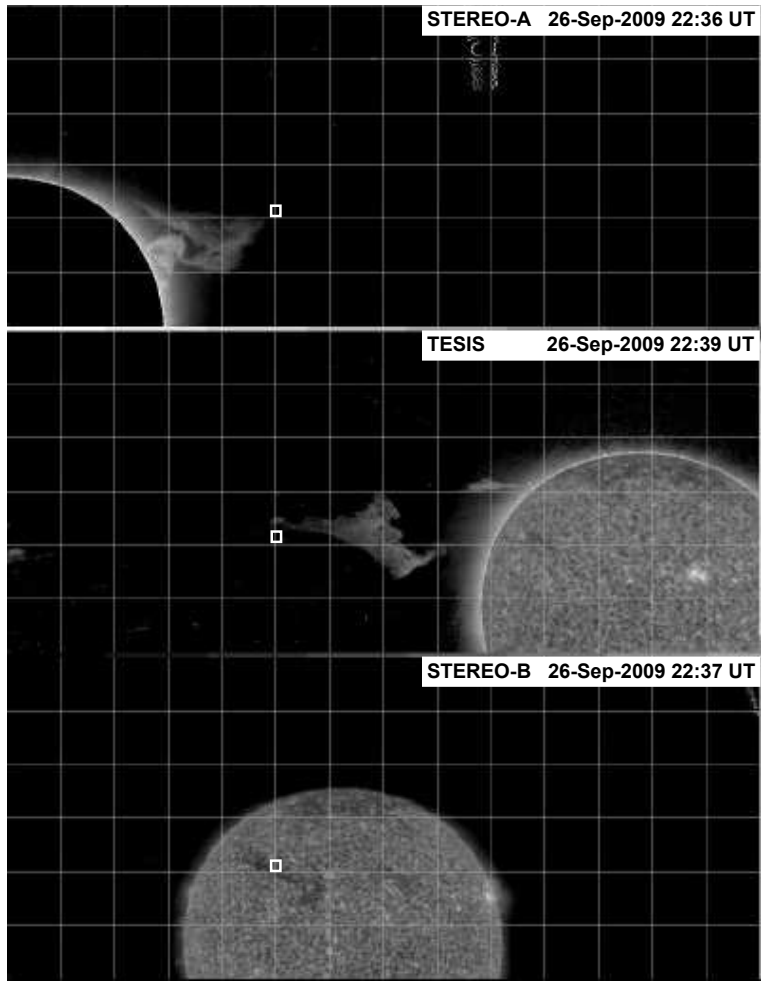}
\end{center}
\caption{ The top, middle, and bottom panels show the three views of the sun
from  the STEREO-A, TESIS and STEREO-B, respectively in Marinus projection. The top part of the
filament is marked by a white box in all images.  }
\label{marinus}
\end{figure}

\begin{figure}      %%%%%%%%%%%%%%%%%%   FIGURE 9
\begin{center}
\includegraphics[angle=0,scale=0.85]{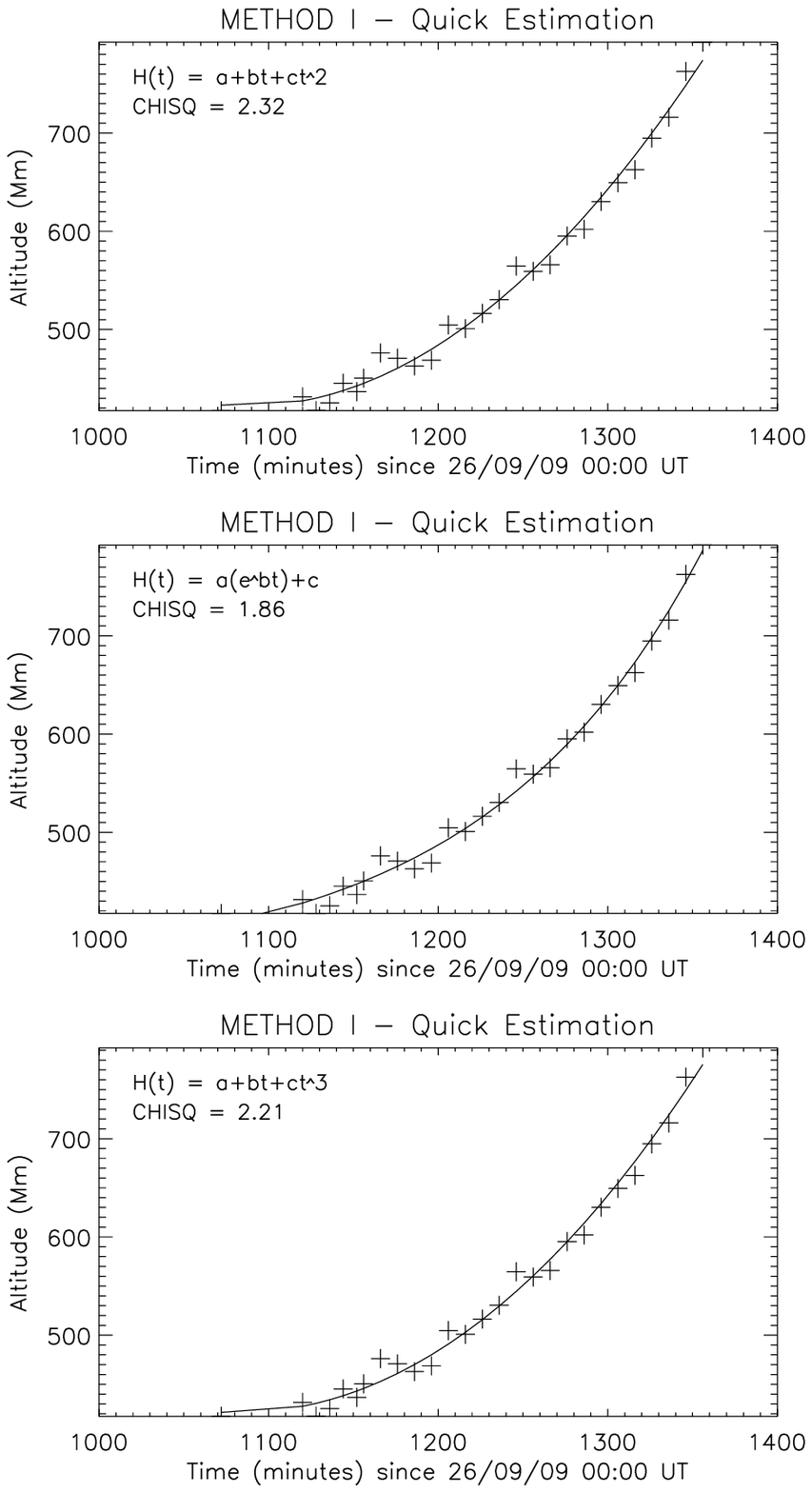}\includegraphics[angle=0,scale=0.85]{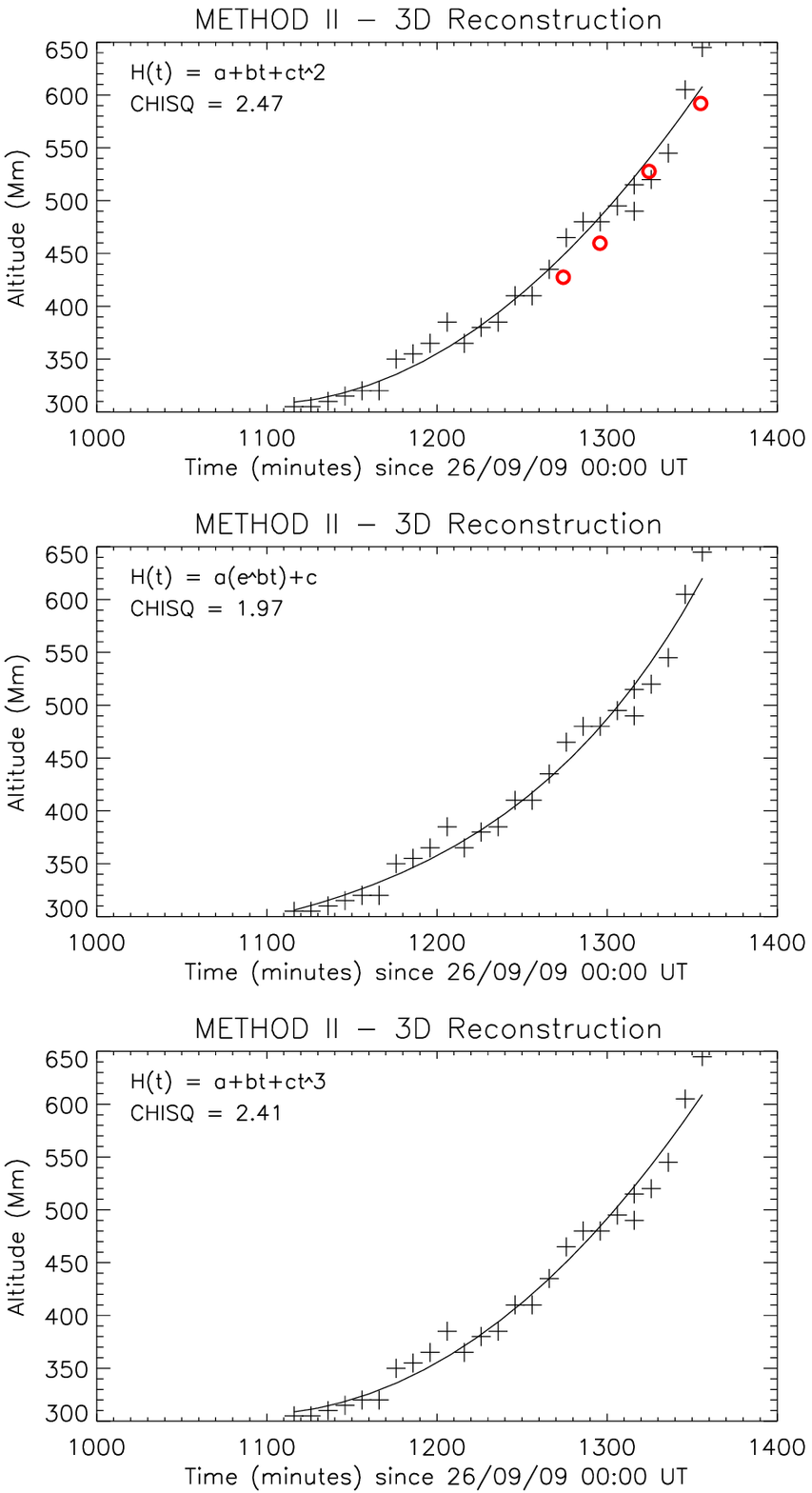}
\end{center}
\caption{The panels on the left and right show the altitude-time curve for the erupting filament derived from simple triangulation method (Section 3.2) and 3-D reconstruction method (Section 3.3). The four data points in red color in top right panel correspond to the altitude reconstructed by using SCC\underline{ }MEASURE method. The altitude-time curve correspond to the rapid acceleration phase of the filament eruption and is fitted for  three functional forms {\it viz.} parabolic, exponential and cubic (from top to bottom).
The reduced chi-square value of the fit is displayed on the top left corner
of each panel. }  \label{artzner-fit}
\end{figure}

The key points in the observational analysis can be summarized as follows:

\begin{enumerate}

\item{The eruption involved two filaments, a large one and a small one, which
were located above the same polarity inversion line, suggesting that they were
embedded in the same, elongated filament channel. The photospheric magnetic
field strengths at the location of the filaments were weak (up to about 3 Gauss).
The two filaments erupted almost simultaneously. In the present analysis we
focused on the eruption of the more prominent large filament.}

\item{We used two different approaches to derive the true  height-time profile of the filament. First we used a simple triangulation method to determine the angle which the filament trajectory makes with respect to the plane-of-sky (POS) and applied correction to the plane-of-sky  height-time profile to derive true height-time profile. Second, we used tomographic approach where we make Marinus projections of the three views of the sun on spheres of radii larger than the solar radii so as to arrive at a common latitude-longitude position of a common feature (filament apex) in all maps. The advantage of the first method is that once we know the propagation angle with respect to POS from triangulation, we can go back and forth in time and correct the POS height-time profile obtained with even one satellite, i.e., durations when only one view is available, for e.g. when in one of the stereoscopic pairs the filament is behind the limb or out of the FOV. However, the method assumes that the propagation angle of the filament with respect to POS does not change substantially over the time of observations.}

\item{During its early rise phase, the filament exhibits the morphology of a twisted sheet. However, its chirality could not be inferred from the images.}

\item{We derived the acceleration and jerk (rate of change of acceleration) profiles for the filament and the CME (Figure~\ref{rapid-accel-profile}). It is believed that the initial rapid acceleration phase, when acceleration is growing, may be suggestive of the physical mechanism behind the eruption (Schrijver et al. 2008).  However, the acceleration curve must be growing and not saturating or slowing down, in other words the jerk should be increasing. By studying the jerk profiles in Figure~\ref{rapid-accel-profile} for the filament and the CME we decided to fit different functional forms to the filament observations only and not to the CME, because the jerk profile of the CME suggests that its acceleration is already saturating. Since the CME observations are available only when it enters the coronagraph's FOV, which is much later than the observations of the filament eruption, we  missed the initial rapid acceleration phase of the CME.}

\item{ We estimate the rapid acceleration phase of the filament between 17:50 UT and 22.33 UT, using the procedure described in section 3.2.1. This phase is marked by two dashed lines in the left panel of Figure~\ref{rapid-accel-profile}. We fit functional forms of a parabolic, exponential and a cubic function to the true  height-time profile of the filament apex during the rapid acceleration phase. These fits to the true height-time curves derived from two independent reconstruction methods described in section 3.2 and 3.3, respectively, are shown in the Figure~\ref{artzner-fit}.}

\end{enumerate}

We now compare our analysis of the eruption kinematics with the predictions
of theoretical eruption models described in the Introduction. We like to note
that conclusions obtained from such a comparison should be read with some care
and not be understood as a way to strictly confirm or rule out certain models.
First, height-time data obtained with current instruments are still not accurate
enough and typically do not have sufficient cadence to allow to pin down
clearly the functional forms of rise profiles, which may behave very similar over
the relatively short time scales of the initial rapid acceleration in solar eruptions.
Also, a clear functional dependence may not be present if several acceleration
mechanisms are at work simultaneously in an eruption. Second, for many models
a proper investigation of the functional dependence of the eruption kinematics
has not yet been reported, and even for most of those for which it was, there
exists no parametric study, which may reveal kinematics of a different functional
dependence than reported for  specific settings of the model parameters.

Our data indicate that the rapid acceleration phase already started before the eruption became visible in the coronagraph data, so we restricted our analysis of the early acceleration phase to the filament observations. Our fits suggest that the filament enters an exponential rise phase at about 17:50 UT, which then appears to saturate from around 22:33 UT. Such exponential initial acceleration is line with many previous studies (see the Introduction) and supports the current picture that both quiescent and active region filament eruptions, and their associated CMEs, are driven by the same mechanisms. It suggests the occurrence of an ideal MHD instability, here most likely the torus instability. We did not find indications of a clear writhing motion of the filament that would suggest the additional occurrence of the helical kink instability, although the twisted appearance of the filament sheet may indicate some untwisting of the magnetic
field during the early phase of the eruption.

The exponential acceleration found here is different from the cases studied by
\citet{schrijver08}, where a cubic (or near-cubic) rise was found for two active
region filament eruptions. However, these authors showed, using numerical simulations,
that a relatively large initial velocity of the erupting structure at the onset of
its rapid acceleration can change the subsequent rise behavior from exponential
to cubic.  The slow rise velocity of the filament (estimated from Figure~\ref{rapid-accel-profile}, plateau in the filament speed curve before the first vertical dashed line) before the filament enters rapid acceleration phase, is relatively small, about $\sim$2.5 km/s, however, comparable to the case described
in \citet{schrijver08}.  The exponential rise also differs from the recent results by \cite{joshi11}, who
found a constant acceleration for both the slow rise and rapid acceleration phases of the
two 3D-reconstructed quiescent prominence eruptions. However, these
authors apparently did not fit functions other than parabolic, also the quality of their fits is not reported.

While the data we considered here support the torus instability as the mechanism
responsible for the initial rapid acceleration of the filament, they do not provide
reasonable clues for the cause its preceding relatively long slow rise phase.
We did not find indications of pre-flare brightening which are often used to draw
conclusions. Hence, we do not find support for tether-cutting or magnetic breakout,
but we cannot rule out the occurrence of these and other reconnection-related
mechanisms, since the magnetic fields in the source region of the eruptions might
have been simply too weak to produce detectable brightening. We therefore refrain
from speculating on the exact underlying mechanism responsible in the present case. However, more studies using the methods developed in this work and encompassing larger sets of observations, including the high-resolution SDO/AIA observations, could provide better clues.

\acknowledgments We thank the anonymous referee for useful comments and suggestions. We thank the SoHO/EIT and STEREO/SECCHI teams
for their open data policy. We thank TESIS/CORONAS team for
providing the data. Financial supports by the European Commission through the FP6
SOLAIRE Network (MTRN-CT-2006-035484) is gratefully acknowledged.

%\bibliographystyle{apj}
%\bibliography{apj-jour,references_stereo}

%%%%%%%%%%%%%%%%%%%% TABLE 1
%
%\begin{table}[ht]
%\caption{Propagation Direction using Simple Triangulation Method } % title of Table
%\centering % used for centering table
%\begin{tabular}{c c c c} % centered columns (4 columns)
%\hline\hline %inserts double horizontal lines
%TESIS & STEREO-A & $\angle\alpha$ & $\angle\beta$ \\ [0.5ex] % inserts table
%%heading
%\hline % inserts single horizontal line
%20:16 UT & 20:16 UT& 41$^\circ$ & 20$^\circ$\\ % insere table
%20:24 UT& 20:26 UT& 41$^\circ$ & 20$^\circ$ \\
%20:48 UT& 20:46 UT& 41$^\circ$ & 20$^\circ$ \\
%20:56 UT& 20:56 UT& 41$^\circ$ & 20$^\circ$ \\
%21:04 UT& 21:06 UT& 41$^\circ$ & 20$^\circ$ \\
%21:59 UT& 21:56 UT& 40$^\circ$ & 21$^\circ$ \\
%22:15 UT& 22:16 UT& 40$^\circ$ & 21$^\circ$ \\ [1ex] % [1ex] adds vertical space
%\hline %inserts single line
%\end{tabular}
%%\footnotetext[1]{Angles $\alpha$ and $\beta$ measured by simple triangulation method.}
%\label{table:propagate} % is used to refer this table in the text
%\end{table}
%

%%%%%%%%%%%%%%%%%%%% TABLE 1
%
\begin{table}[ht]
\caption{Propagation Direction using Simple Triangulation and Marinus Method } % title of Table
\centering % used for centering table
\begin{tabular}{c c c c c c} % centered columns (4 columns)
\hline\hline %inserts double horizontal lines
TESIS & STEREO-A & $\angle\alpha$\footnotemark[1] & $\angle\beta$\footnotemark[1]& Latitude\footnotemark[2] & Longitude\footnotemark[2]\\ [0.5ex] % inserts table
%heading
\hline % inserts single horizontal line
20:16 UT & 20:16 UT& 41$^\circ$ & 20$^\circ$ & 28$^\circ$&148$^\circ$\\ % insere table
20:24 UT& 20:26 UT& 41$^\circ$ & 20$^\circ$ & 28$^\circ$ &148$^\circ$\\
20:48 UT& 20:46 UT& 41$^\circ$ & 20$^\circ$ & 27$^\circ$ &149$^\circ$\\
20:56 UT& 20:56 UT& 41$^\circ$ & 20$^\circ$ & 25$^\circ$ &149$^\circ$\\
21:04 UT& 21:06 UT& 41$^\circ$ & 20$^\circ$ & 25$^\circ$ &149$^\circ$\\
21:59 UT& 21:56 UT& 40$^\circ$ & 21$^\circ$ & 21$^\circ$ &152$^\circ$\\
22:15 UT& 22:16 UT& 40$^\circ$ & 21$^\circ$ & 19$^\circ$&152$^\circ$\\ [1ex] % [1ex] adds vertical space
\hline %inserts single line
\end{tabular}
\footnotetext[1]{Angles $\alpha$ and $\beta$ measured by simple triangulation method.}
\footnotetext[2]{Carrington latitude and longitude using Marinus method.}
\label{table:propagate} % is used to refer this table in the text
\end{table}

\end{document}